\documentclass[pdflatex,sn-mathphys-num]{sn-jnl} 
\usepackage{graphicx}
\usepackage{multirow}
\usepackage{amsmath,amssymb,amsfonts}
\usepackage{amsthm}
\usepackage{mathrsfs}
\usepackage[title]{appendix}
\usepackage{xcolor}
\usepackage{textcomp}
\usepackage{float}
\usepackage{manyfoot}
\usepackage{booktabs}
\usepackage{algorithm}
\usepackage{algorithmicx}
\usepackage{algpseudocode}
\usepackage{listings}

\theoremstyle{thmstyleone}%

\theoremstyle{thmstyletwo}

\theoremstyle{thmstylethree}%

\begin{document}

\title[Article Title]{Comparative Analysis of Black Hole Mass Estimation in Type-2 AGNs: Classical vs. Quantum Machine Learning and Deep Learning Approaches}

\author[1]{\fnm{Sathwik} \sur{Narkedimilli}}\email{21bcs103@iiitdwd.ac.in}

\author[1]{\fnm{Venkata Sriram} \sur{Amballa}}\email{21bcs008@iiitdwd.ac.in}

\author[1]{\fnm{N V Saran} \sur{Kumar}}\email{21bcs075@iiitdwd.ac.in}

\author[1]{\fnm{R Arun} \sur{Kumar}}\email{21bcs093@iiitdwd.ac.in}

\author[1]{\fnm{R Praneeth} \sur{Reddy}}\email{21bcs090@iiitdwd.ac.in}

\author[2]{\fnm{Satvik} \sur{Raghav}}\email{satvikraghav007@gmail.com}

\author[3]{\fnm{Manish M}}\email{mmodani@nvidia.com}

\author*[4]{\fnm{Aswath Babu} \sur{H}}\email{aswath@iiitdwd.ac.in}

\affil[1]{\orgdiv{Department of Computer Science}, \orgname{Indian Institute of Information Technology (IIIT) Dharwad}, \orgaddress{\city{Dharwad}, \state{Karnataka}, \country{India}}}

\affil[2]{\orgdiv{Department of Electronics and Communication Engineering}, \orgname{Amrita School of Engineering}, \orgaddress{\city{Bengaluru}, \state{Karnataka}, \country{India}}}

\affil[3]{\orgname{NVIDIA}, \orgaddress{\city{Pune}, \country{India}}}

\affil[4]{\orgdiv{Department of Humanities and Science}, \orgname{Indian Institute of Information Technology (IIIT) Dharwad}, \orgaddress{\city{Dharwad}, \state{Karnataka}, \country{India}}}

\abstract{
In the case of Type-2 AGNs, estimating the mass of the black hole is challenging. Understanding how galaxies form and evolve requires considerable insight into the mass of black holes. This work compared different classical and quantum machine learning (QML) algorithms for black hole mass estimation, wherein the classical algorithms are Linear Regression, XGBoost Regression, Random Forest Regressor, Support Vector Regressor (SVR), Lasso Regression, Ridge Regression, Elastic Net Regression, Bayesian Regression, Decision Tree Regressor, Gradient Booster Regressor, Classical Neural Networks, Gated Recurrent Unit (GRU), LSTM, Deep Residual Networks (ResNets) and Transformer-Based Regression. On the other hand, quantum algorithms including Hybrid Quantum Neural Networks (QNN), Quantum Long Short-Term Memory (Q-LSTM), Sampler-QNN, Estimator-QNN, Variational Quantum Regressor (VQR), Quantum Linear Regression(Q-LR), QML with JAX optimization were also tested. The results revealed that classical algorithms gave better R², MAE, MSE, and RMSE results than the quantum models. Among the classical models, LSTM has the best result with an accuracy of 99.77\%. Estimator-QNN has the highest accuracy for quantum algorithms with an MSE of 0.0124 and an accuracy of 99.75\%. This study ascertains both the strengths and weaknesses of the classical and the quantum approaches. As far as our knowledge goes, this work could pave the way for the future application of quantum algorithms in astrophysical data analysis.
}


\maketitle

\section{Introduction}\label{sec1}

Active Galactic Nuclei (AGNs) involve the most powerful phenomena in the Universe and they are known to be highly luminous objects. There exists a supermassive black hole at their center. One of the most important steps towards understanding galaxy formation and evolution lies in accurately measuring black-hole mass in AGNs. In particular, the scaling relations between black hole mass and properties of the host galaxy can help to unlock the black hole and host galaxy co-evolutionary paths throughout cosmic time \cite{Ferrarese_2000}. However, these scaling relations are hard to pin down for some types of AGNs, because the accretion disk, specifically, the broad-line region (BLR), is obscured in the spectrum. An example of such an AGN is a Type-2 AGN, which we addressed through our study here.

The dusty torus surrounding the black hole in a type-2 AGN \cite{Netzer_2015} means that one can observe only the narrow-emission lines from the narrow-line region (NLR), and standard methods such as reverberation mapping (which rely on the broad lines from type-1 AGNs, which are not visible) can’t be applied in type-2 systems. In our literature survey, we’ve noticed that the black hole mass can still be estimated for type-2 systems using scaling relations based on narrow-line properties, which gives us an indirect handle on the mass in obscured systems. Knowing the black hole mass, and therefore the accretion rate for these different classes of AGNs is crucial for probing the general relation between the growth in the black hole and the evolution of the host galaxy~\cite{Zakamska_2003}.

Here, we present the implementation of classical and quantum machine learning algorithms to estimate black hole mass in Type-2 AGNs. Classical approaches in machine learning, including linear regression, XGBoost, random forest regression, and LSTM algorithms, are widely used in different domains of science, including astrophysics, to model complex relationships between input features and the output of predictions. In contrast, quantum machine learning (QML) approaches, including Hybrid Quantum Neural Networks, Q-LSTM, and Estimator-QNN algorithms, promise quantum computational advantages in solving certain black hole mass estimation tasks leading to a new paradigm.

Using this comparison, we can establish the effectiveness of classical ML and quantum algorithms in predicting the mass of black holes in the AGNs of Type-2. We gauge the performance of ML algorithms in terms of R², MAE, MSE, and RMSE. We found that classical algorithms performed slightly better in all three metrics than their corresponding quantum algorithms. The classical LSTM achieved the best value regarding accuracy and all error metrics. However, quantum algorithms such as Estimator-QNN performed almost as competitively, hinting at the vast potential of quantum computation in the future. This study presents a detailed comparison of both classical and quantum techniques that can be used for machine learning on astrophysical data. It provides a reference for future applications of machine learning in astrophysics.

In this study, quantum machine learning algorithms performed nearly on par with classical machine learning and deep learning approaches. However, as quantum hardware continues to advance—with enhanced error correction and noise reduction techniques—it is anticipated that quantum ML will soon surpass existing classical methods, potentially revolutionizing astrophysical data analysis.

Motivated by the challenge of deciphering black hole-host galaxy evolution, our study bridges observational gaps in Type-2 AGNs—where obscured broad-line regions limit traditional methods—by leveraging scaling relations based on narrow emission lines. We comprehensively compare classical algorithms (such as LSTM, Random Forest, and XGBoost) with emerging quantum techniques (like Estimator-QNN and Q-LSTM), aiming not only to refine black hole mass estimation in obscured systems but also to unlock the potential of quantum computing in astrophysical data analysis.

The paper is organized with the background Section-\ref{sec2} and a review of relevant literature Section-\ref{sec3} to establish context and identify gaps. Section-\ref{sec4} outlines the methodology, followed by results in Section-\ref{sec5}. The discussion in Section-\ref{sec6} connects findings to broader implications. The conclusion Section-\ref{sec7} summarizes contributions, and Section-\ref{sec8} highlights future research directions.

\section{Background}\label{sec2}

\subsection{Active Galactic Nuclei (AGNs)}\label{subsec2}

Active Galactic Nuclei (AGNs) \cite{Perlman2013} are powerful and compact regions found at the centers of certain galaxies, where supermassive black holes are accreting vast amounts of matter. This accretion process triggers intense electromagnetic radiation, making AGNs some of the brightest objects in the universe. AGNs are classified into several types based on their observational properties, including Type-1 and Type-2 AGNs, where the difference is attributed to the angle at which the AGN is viewed relative to its obscuring material (often referred to as the torus). Type-2 AGNs, specifically, have their central engines obscured from direct view, meaning their broad emission lines are hidden by dust and gas, making it more challenging to estimate black hole masses accurately.

AGNs are critical to understanding galaxy formation and evolution, as they regulate star formation and influence their host galaxies through various feedback processes. These processes involve the release of energy and momentum—via radiation, jets, and winds—that interact with the surrounding interstellar medium, thereby either quenching or triggering star formation and reshaping the galaxy’s structure. AGNs exhibit complex behaviors across the entire electromagnetic spectrum, from radio to gamma rays. Their energy output and variability offer key insights into the physics of accretion and the dynamic interplay between black holes and their environments. Investigating AGNs, particularly Type-2 AGNs, is therefore essential for advancing our understanding of supermassive black hole growth and their broader cosmological impact.

\subsection{Traditional Methods for Black Hole Mass Estimation}\label{subsec2}

Estimating the mass of supermassive black holes in Active Galactic Nuclei (AGNs) has traditionally been achieved using techniques such as reverberation mapping \cite{CACKETT2021102557} and the $M-\sigma$ relation. Reverberation mapping involves observing the time lag between variations in the AGN’s continuum emission and the corresponding changes in its broad emission lines. This time lag is used to estimate the size of the broad-line region (BLR), which, combined with the velocity of the emitting gas, provides an estimate of the black hole’s mass via the virial theorem. While effective for Type-1 AGNs, where the broad-line region is visible, on the other hand it is not directly applicable to Type-2 AGNs due to the obscured central region.

Another widely used method is the $M-\sigma$ relation, which correlates the mass of a galaxy's supermassive black hole with the velocity dispersion of its stars in the bulge. This empirical relation has been crucial in estimating black hole masses in both active and inactive galaxies. However, this method is also subject to limitations, such as intrinsic scatter in the relation and observational biases, especially for galaxies where the stellar velocity dispersion is difficult to measure accurately. These traditional methods provide valuable insights but often struggle to capture the complexity of black hole environments, particularly in AGNs with obscured central engines like Type-2 AGNs~ \cite{Ferrarese_2000}.

\subsection{Machine Learning and Deep Learning in Astrophysics}\label{subsec2}

In recent years, machine learning (ML) and deep learning (DL) have revolutionized data-driven research across various fields, including astrophysics. These techniques offer the ability to analyze vast amounts of data efficiently, identifying patterns and relationships that may be difficult to discern through traditional statistical methods \cite{CACKETT2021102557}. In astrophysics, ML and DL algorithms have been applied to tasks such as galaxy classification, star formation rate estimation, and even the detection of exoplanets. When applied to black hole mass estimation, these methods offer a way to build models that can process large astronomical datasets, learning from observed features and improving mass prediction accuracy.

Deep learning, in particular, with its multi-layered neural networks \cite{Meher2021}, has been successful in recognizing complex patterns in high-dimensional datasets, making it ideal for handling the intricate and noisy data that often arise in astronomical surveys. Moreover, ML models can be trained to generalize from incomplete or biased datasets, offering a potential solution to the limitations of traditional methods. Using machine learning to complement or replace classical approaches, astronomers can refine black hole mass estimates, especially in challenging cases like Type-2 AGNs, where observational data is incomplete or obscured.

\subsection{Quantum Machine Learning (QML)}\label{subsec2}

Quantum Machine Learning (QML) \cite{Biamonte2017} combines the principles of quantum computing with machine learning techniques to harness the power of quantum mechanics to data processing. Quantum computers can process and analyze vast amounts of data in ways that classical computers cannot, thanks to phenomena such as superposition and entanglement. QML algorithms can explore large feature spaces more efficiently than classical ML, providing an edge in tackling complex problems in astrophysics. The combination of quantum parallelism and ML models allows QML to address high-dimensional data and optimize calculations that would be infeasible using traditional methods.

The potential of QML in astrophysics lies in its ability to perform calculations faster and more accurately, particularly in scenarios where the data is vast, noisy, and multi-dimensional. For black hole mass estimation, QML can process large astronomical datasets in parallel, identifying subtle correlations and patterns that classical algorithms might miss. As quantum computing advances, its application in machine learning will likely bring transformative breakthroughs in astrophysics, enabling researchers to push the boundaries of knowledge in fields like black hole physics and cosmology.

\section{Literature Survey}\label{sec3}

In this section, we discuss the existing literature on black hole mass estimation which encompasses a spectrum of methodologies, including empirical scaling relations, reverberation mapping, variability-based techniques, and classical machine learning approaches. Empirical scaling relations and reverberation mapping have laid the groundwork by using observable correlations and time-delay measurements, despite inherent uncertainties and limitations from indirect data. Variability-based methods provide dynamic insights, particularly for blazars, while classical machine learning techniques have enhanced estimation efficiency by leveraging extensive datasets. This survey synthesizes these diverse approaches and establishes their relevance as a foundation for our study, which explores how quantum machine learning could overcome current challenges and advance the precision of black hole mass estimation.

Empirical scaling relations and reverberation mapping are key methods for estimating black hole masses in AGNs using emission lines and light variability, though they rely on indirect measurements and inherent assumptions. Below are the studies that illustrate these techniques.

The study titled ``Black hole mass estimation for active galactic nuclei from a new angle" \cite{euifhibibsibgiwbegiwbegib} introduces a method that relies on a scaling relation between narrow [O~III] and broad H$\alpha$ emission lines, enabling black hole mass estimation in type 2 AGNs where broad emission lines are obscured. This empirical approach allows estimation in galaxies previously inaccessible to traditional methods, albeit with reliance on high-quality spectroscopic data and assumptions on emission line consistency. Using this method, black hole masses were estimated for a sample of approximately 10,000 type 2 AGNs from the Sloan Digital Sky Survey (SDSS). The derived M${BH}$–$\sigma$ and M${BH}$–M* scaling relations were consistent with those observed in type 1 AGNs and inactive galaxies, validating the reliability of the approach.

The research study titled ``Measuring the Masses of Supermassive Black Holes"\cite{peterson2015measuring} examines the techniques for black hole mass measurement, focusing on reverberation mapping and its application to accreting supermassive black holes. The study emphasizes the importance of scaling relationships derived from reverberation mapping, which, despite their utility, face challenges related to reliance on indirect measures and assumptions that introduce uncertainties in mass estimates.  Reverberation mapping methods were found to yield black hole mass estimates with typical uncertainties of 0.3–0.4 dex, depending on the quality of data and underlying assumptions.

The study entitled ``Black Hole Masses and Methods to Estimate the Mass" \cite{xie2005agn} proposes a new technique to estimate black hole masses in blazars using variability timescales in blazar emissions as a proxy for mass. This method leverages high-energy observational data, allowing more efficient mass estimation. However, its limitations include the assumptions that variability arises solely from processes near the black hole and sensitivity to noise and observational quality, which can affect its reliability. The study demonstrated the feasibility of the variability-based mass estimation method, achieving accuracy comparable to traditional techniques for a subset of well-monitored blazars.

The studies below highlight the use of Machine Learning and Deep Learning techniques in black hole mass estimation.

The paper titled ``AGNet: Weighing Black Holes with Machine Learning" \cite{lin2020agnet} proposes AGNet, a neural network model using quasar light curves for supermassive black hole (SMBH) mass estimation. This deep learning framework reduces the costs associated with mass estimation and produces results comparable to traditional methods. Despite this advantage, the model’s accuracy is limited by the quality of the input data and uncertainties in mass estimation from light curves, highlighting the challenges of machine learning applications in astrophysics. AGNet achieved a median deviation of less than 0.4 dex in SMBH mass estimates, showing promise as an alternative to spectroscopic methods in cases with high-quality light curve data.

The work titled ``Predicting the black hole mass and correlations in X-ray reverberating AGNs using neural networks" \cite{chainakun2022predicting} employs neural networks to estimate black hole masses based on X-ray reverberation data. This approach aims to overcome challenges related to insufficient temporal resolution and noise that affect traditional reverberation mapping. Despite showing promise, the proposed solution still requires extensive high-quality training data, making it sensitive to noise and observational inconsistencies. The study reported an average relative error of 10\% in black hole mass predictions, demonstrating the potential of neural networks in improving X-ray reverberation-based techniques.

The study titled ``Interpretable Machine Learning for Finding Intermediate-mass Black Holes" \cite{pasquato2024interpretable} utilizes an interpretable machine learning framework to identify intermediate-mass black holes (IMBHs), which are difficult to detect due to their ambiguous observational features. The approach enhances transparency in identifying IMBH candidates based on features like X-ray luminosity and variability. However, the model's dependency on well-characterized datasets limits its generalizability, posing challenges for robust IMBH detection. The model successfully identified 27 IMBH candidates, with an estimated detection efficiency of 85\% based on simulated datasets.

Moreover, a study titled ``Machine Learning in Astrophysics and Cosmology"\cite{casas2023machine} presents methodologies involving convolutional neural networks to analyze the cosmic microwave background (CMB). This approach uses neural networks trained on realistic simulations to detect radio galaxies and analyze the CMB properties, aiming to improve traditional analysis techniques. However, the reliance on accurate simulations and difficulties in generalizing to complex observational data are significant challenges faced by this machine learning-based solution. The study achieved a classification accuracy of 93\% for radio galaxies and demonstrated improved noise reduction in CMB analyses compared to traditional methods.

The article titled ``Surveying the Reach and Maturity of Machine Learning and Artificial Intelligence in Astronomy`` \cite{fluke2020surveying} reviews the current state of ML and AI in solving astrophysical problems. The paper highlights gaps in the integration of AI techniques into astronomy and provides a framework to evaluate the effectiveness of ML/AI. Despite significant progress, challenges such as data quality, model interpretability, and the need for domain expertise remain key limitations to the broad application of AI in astronomy. The survey revealed that ML models achieved an average accuracy improvement of 15–20\% over traditional methods in classification tasks but underscored the need for more interpretable and reliable models.

The paper titled ``Predicting Supermassive Black Hole Mass with Machine Learning Methods"\cite{he2022predicting} proposes regression-based machine learning models for predicting supermassive black hole masses. By addressing the challenges of traditional empirical scaling relations, the authors aim to develop an efficient and accurate estimation framework. However, the need for high-quality data and sensitivity to training datasets present obstacles in effectively generalizing the models across varied galactic environments. The regression models achieved an R² of 0.87 in predicting SMBH masses, indicating strong performance under controlled data conditions.

Despite significant progress with empirical scaling relations, reverberation mapping Variability-Based Techniques, and classical machine learning techniques, notable challenges remain. Many of these methods depend on high-quality, often scarce observational data, suffer from uncertainties due to indirect measurements and noise, and face difficulties in model interpretability and generalizability across diverse AGN populations. Quantum machine learning presents a promising frontier that could potentially address these gaps by leveraging quantum computational advantages. By integrating quantum algorithms with advanced machine learning, future research may overcome current limitations, improve data processing efficiency, and enhance the accuracy and reliability of black hole mass estimations. The study below outlines the advantages that quantum computing offers.

The paper titled ``Recent Advances in Quantum Machine Learning"\cite{zhang2020recent} explores methods to enhance computational efficiency in quantum systems by integrating quantum algorithms with traditional simulations. While quantum-enhanced simulations hold promise, they are currently constrained by significant challenges such as hardware limitations and scalability issues, limiting the practical application of these techniques. Quantum algorithms demonstrated speedups of up to 40\% in small-scale simulations, showcasing their potential for scaling astrophysical computations in the future.

\section{Methodology}\label{sec4}

\subsection{Hardware and Computing tools used}

For the execution of the classical algorithms, including Classical Deep Learning Algorithms, Classical Neural Networks, GRU (Gated Recurrent Unit), Deep Residual Networks (ResNets), Transformer-Based Regression, LSTM, as well as various classical regression models like Linear Regression, XGBoost, Random Forest Regressor, SVR, Lasso Regression, Ridge Regression, Elastic Net Regression, Bayesian Regression, Decision Tree Regressor, and Gradient Booster Regressor, we utilized T4 Graphics Processing Units (GPUs) and TPUs procured through a cloud-based environment. These T4 GPUs and TPUs offered high computational efficiency, significantly reducing the computing time for the model training and evaluation processes, and enabling more efficient experimentation and faster iterations in our research.

Hybrid QNN with CUDA-Quantum was executed for the quantum algorithms using NVIDIA GPUs for high-performance computations. For Sampler-QNN, Estimator-QNN, VQR, and QML with JAX optimization, we utilized the AER simulator to efficiently simulate the quantum machine learning algorithms offered by Qiskit. Additionally, for Q-LSTM and Q-LR, CuQuantum-enabled GPUs were employed to accelerate quantum instances, ensuring optimized performance for these quantum algorithms using Pennylane in the study.

\subsection{Dataset}\label{subsec4}

\begin{figure}[ht]
    \centering
    \includegraphics[width=1\textwidth]{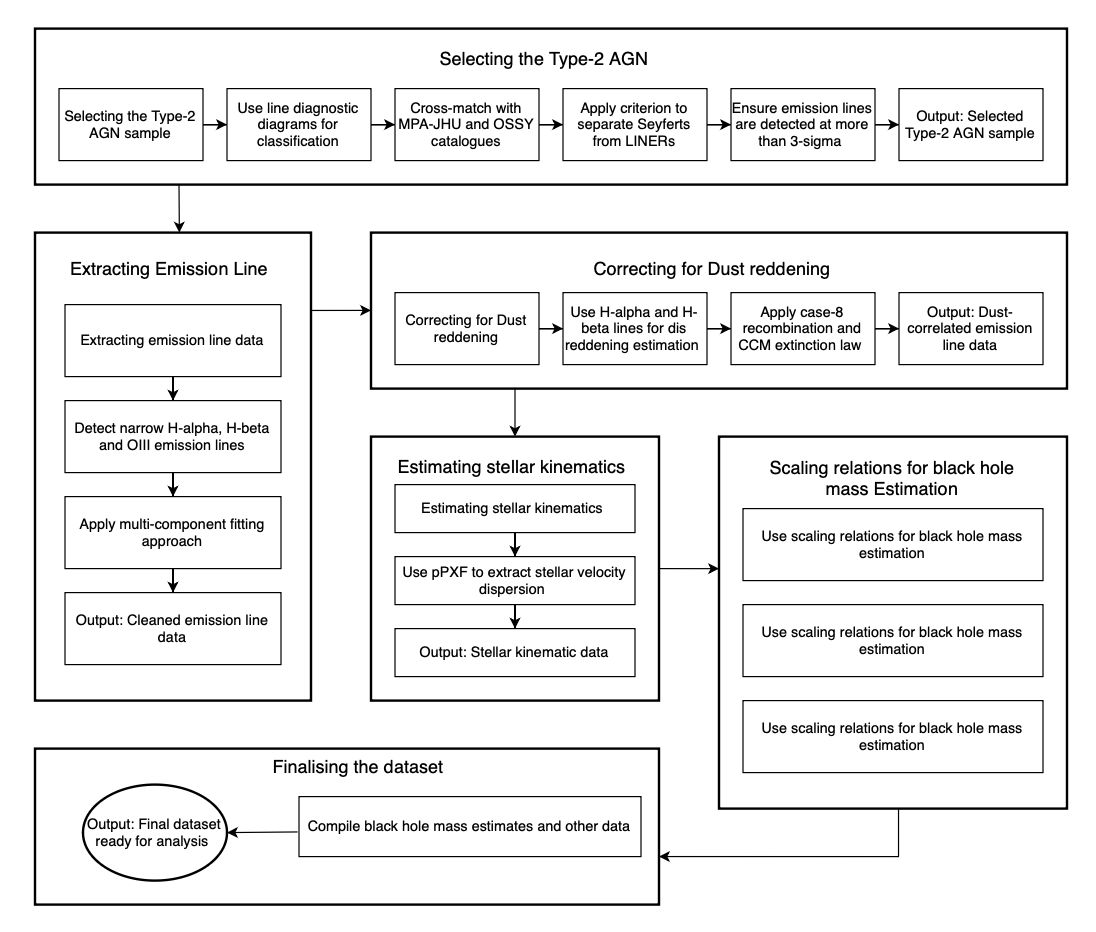}
    \caption{Steps for Blackhole Mass Estimation Catalog Creation}
    \label{fig:sample_image33}
\end{figure}

We obtained the required dataset by following the methods explained in Dalya {\em et.al}  titled ``Black hole mass estimation for Active Galactic Nuclei from a new angle" \cite{euifhibibsibgiwbegiwbegib}. We have sourced the dataset from the Sloan Digital Sky Survey (SDSS) Data Release 7 (DR7) which focuses on approximately 10,000 Type-2 Active Galactic Nuclei (AGN) spectra. These AGNs are obscured by a dusty torus, preventing direct observation of the BLR. Since black hole mass estimation relies on narrow emission line measurements such as [OIII], H$\alpha$, and H$\beta$, additional features, including stellar velocity dispersion ($\sigma_*$) and stellar mass estimates, are used to support the analysis. The dataset includes critical spectral features necessary for applying scaling relations to estimate black hole masses. Pre-processing steps ensure that the data quality is sufficient for accurate measurements, making it ideal for comparing classical and quantum machine learning models in black hole mass estimation. 
Fig.\ref{fig:sample_image33} describes how the required catalog is obtained.

\subsection{Data Pre-Processing}\label{subsec4}

In the data preprocessing workflow, we begin by mounting the dataset from the given path and handling missing values through mean imputation, ensuring no missing data points that affect the model’s accuracy. The features and the target variable, which in this case is the logarithmic black hole mass (log\_bh\_mass), are defined next. To tackle high dimensionality, Principal Component Analysis (PCA) is applied to reduce the number of features, allowing the model to focus on the most relevant ones. The features and target variable are then normalized using either Min-Max scaling or standard scaling to bring them within a consistent range, ensuring that the model is not influenced by the varying magnitudes of different features. Finally, the data is split into training and testing sets in an 80:20 ratio to assess the model’s performance on unseen data.

Additional preprocessing steps involve feature selection to reduce multi-collinearity, which helps eliminate redundant features and make the dataset easier to interpret. Data reshaping is also crucial, especially for sequential models like Long Short-Term Memory (LSTM), where the input format must be adjusted appropriately. Overall, this systematic workflow ensures the dataset is clean, relevant, and well-prepared for machine learning models, facilitating better model performance and interpretability.

\subsection{Feature Engineering}\label{subsec4}

In this study, a comprehensive set of features is selected for black hole mass estimation in Type-2 AGNs. The feature engineering is carried out using the LDA plots and the random forest feature importance plots.  The chosen features encompass key spectral properties and physical attributes of the AGNs and their host galaxies. Specifically, spectral line fluxes and their associated errors, such as the H$\beta$ flux, [OIII] 5007 flux, H$\alpha$ flux, and [NII] 6584 flux, are utilized. These emission lines are critical for understanding the kinematics and ionization states within the AGNs. Additionally, photometric data from different bands, including the PSF magnitudes (u, g, r, i, z) and their respective errors, contribute to capturing the overall luminosity and color profile of the AGN. Stellar velocity dispersion, represented by the logarithm of stellar sigma, is also included as a key feature indicative of the central black hole’s influence on surrounding stars.

In addition to spectral and photometric data, several stellar mass estimates and morphological parameters are integrated into the feature set. Stellar mass estimates, derived from broadband spectral energy distribution (SED) fitting, include median values along with 16th and 84th percentile uncertainties for total stellar mass (logMt), bulge mass (logMb), and disk mass (logMd). These provide insight into the host galaxy's structure. Morphological features such as the bulge-to-total ratio (b/t) and half-light radius in both the g and r bands, along with their errors, further describe the host galaxy's geometry. Together, these features enable the model to analyze both the physical and spectral characteristics of the AGNs, ensuring a well-rounded input for the machine learning, Deep Learning, and QML algorithms used to estimate the black hole mass (log\_bh\_mass).

\subsection{Classical ML and DL Workflow}\label{subsec4}

The workflow for executing classical ML and DL algorithms for black hole mass estimation can be broadly divided into three key stages: modeling, training, and evaluation. In the modeling stage, various classical and deep learning models are applied to the dataset. The deep learning models include Classical Neural Networks, ResNet, GRU, LSTM, and transformer-based regression models, each designed to handle different types of data complexities. On the other hand, the classical ML models, such as Bayesian regression, decision tree, elastic net regression, and support vector regression (SVR), aim to extract relevant patterns from the data using traditional regression techniques.

For the deep learning models, the Classical Neural Networks are structured as feed-forward networks with multiple input, hidden, and output layers. These networks use activation functions like ReLU and sigmoid to learn non-linear relationships in the data. ResNet introduces skip connections and residual blocks to address the vanishing gradient problem, allowing the model to retain important information across layers. In time-series data modeling, GRU and LSTM excel by capturing sequential dependencies, with GRU optimizing recurrent connections and LSTM specializing in capturing long-term dependencies. Finally, Transformer-based regression uses a self-attention mechanism, adapted to regression tasks, to model complex relationships between input features.

\bigskip
\begin{figure}[ht]
    \centering
    \includegraphics[width=1\textwidth]{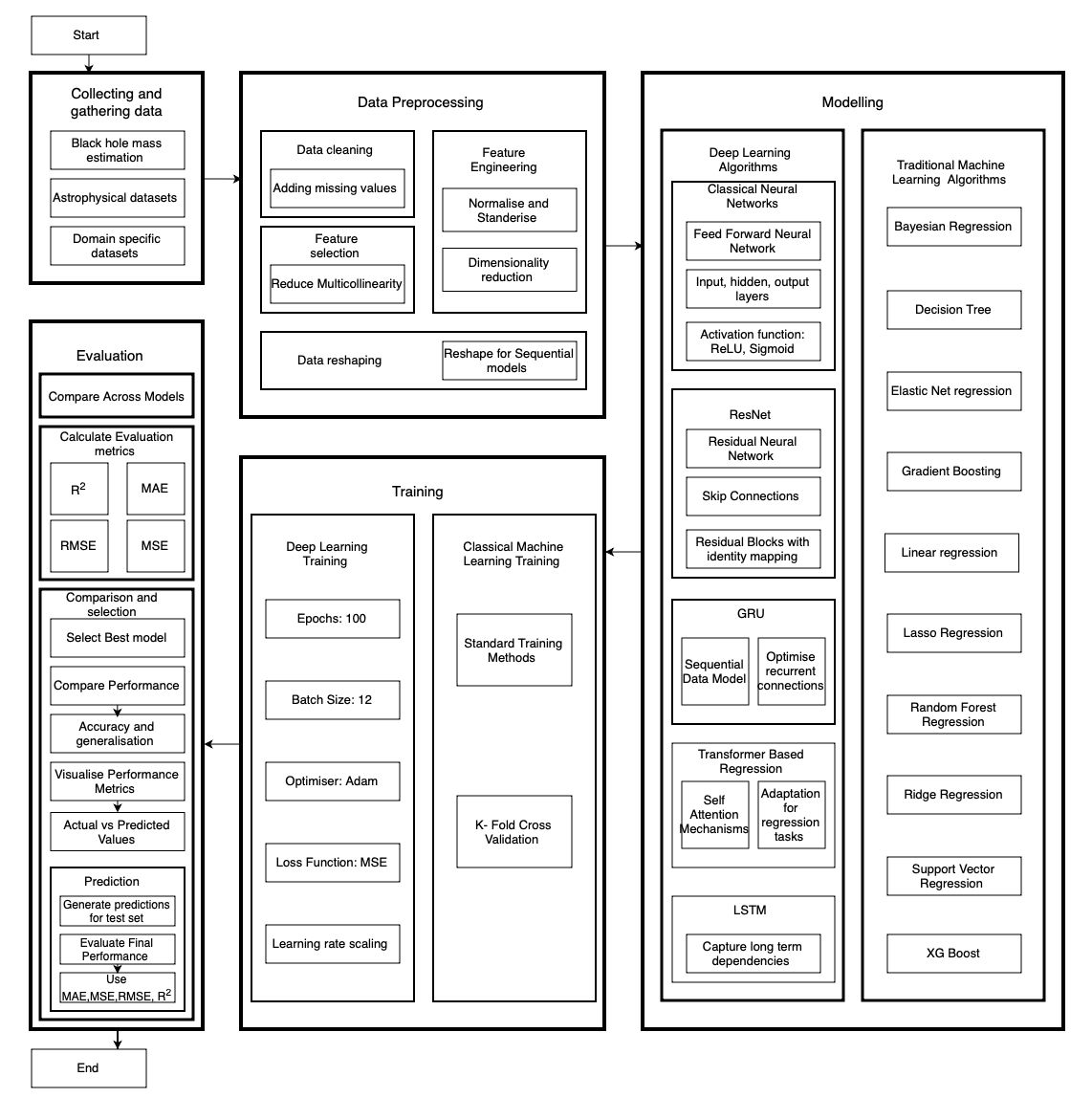}
    \caption{Workflow diagram for Executing Classical DL and ML Algorithms}
    \label{fig:sample_image3333424}
\end{figure}
On the classical machine learning front, algorithms like Bayesian regression, decision trees, elastic net regression, gradient boosting, and XGBoost are applied. These methods are well-suited for tabular data and work efficiently in extracting patterns and relationships between the astrophysical features and black hole mass. Models like random forest regression and support vector regression are particularly useful in handling high-dimensional data, while regularization techniques like lasso and ridge regression help to avoid overfitting by penalizing large coefficients in linear models.

In the training stage, the deep learning models are trained over 100 epochs, with a batch size of 32, using the Adam optimizer. The mean squared error (MSE) is used as the loss function to minimize the error between predicted and actual black hole mass values. A learning rate scheduler adjusts the learning rate to optimize convergence. For classical ML models, standard training methods are employed, often involving grid search for hyper-parameter tuning to find the best configuration for each model.

The evaluation stage involves calculating common metrics, such as R², MAE, RMSE, and MSE, across all models to gauge their performance. These metrics are compared to determine which model or algorithm is best suited for black hole mass estimation in Type-2 AGNs.

Fig.\ref{fig:sample_image3333424} below describes the workflow for executing the classical ML and DL algorithms for black hole mass estimation.

\subsection{QML Workflow}\label{subsec4}

The QML workflow for black hole mass estimation begins with Model Execution, where various quantum and hybrid models are employed. In the Quantum Processing Setup, the process begins with the design of the quantum circuit, which is a critical component of quantum machine learning models. The Quantum Circuit Design employs a ZZFeatureMap to encode classical input data into the quantum circuit. This feature map is particularly useful for capturing correlations between features in the dataset. The circuit is designed to use 4 qubits, which provide the quantum computational resources needed to process the data effectively. To execute the quantum operations, the setup integrates quantum platforms such as Pennylane, Qiskit, and CUDA-Quantum (cuda-Q), ensuring compatibility and efficiency in running quantum algorithms on different platforms. This setup forms the foundation for the subsequent quantum computations in the workflow.

The first model in this sequence is the Hybrid QNN using cuda-Q. This model combines classical and quantum computing by initializing a quantum circuit that implements rotation gates like RY and RX to process input features. Classical layers, defined in PyTorch, are integrated with fully connected and dropout layers to enhance model performance and prevent overfitting. During the forward pass, the output from the classical layers is passed through a quantum layer to incorporate quantum computational advantages. Finally, the optimization step employs the Adam optimizer to minimize the Mean Squared Error Loss (MSELoss), ensuring that the model can effectively learn and generalize from the data.

Next, the Quantum Linear Regression (QLR) model starts by utilizing ZZFeatureMap for feature mapping, where data is embedded using RX and RY rotation gates. This embedding allows the quantum circuit to process classical input data in a quantum-enhanced manner. Afterward, the model defines a quantum ansatz, which is essentially a parametrized quantum circuit, to perform linear regression with the quantum-mapped features. To ensure robustness and prevent overfitting, 5-fold cross-validation is applied, splitting the data into five sets for iterative training and evaluation. Optimization is handled using the GradientDescentOptimizer, which minimizes the MSE, refining the model's predictions with each iteration.

The Q-LSTM model leverages both classical and quantum components to handle sequential data for time-series predictions. In this model, the classical LSTM layers are first defined in PyTorch, handling the temporal dependencies in the input data. These layers are then augmented by quantum blocks, where quantum gates implemented via QNode are applied within the LSTM architecture. The forward pass in this model involves feeding the input into a quantum memory unit, enabling quantum-enhanced handling of sequence data. The Adam optimizer, coupled with MSELoss, is employed to minimize the prediction error during the training process, optimizing both the classical and quantum components for accurate time-series predictions.

The Estimator QNN utilizes a quantum circuit with 4 or 5 qubits to process input data, which is then combined with classical layers. The classical layers are built to concatenate outputs from both the quantum and classical components of the model. This hybrid approach allows for enhanced processing power by leveraging both quantum and classical resources. During the forward pass, the model combines outputs from these layers to generate predictions. Optimization is handled by the Adam optimizer, with MSELoss as the error metric, ensuring the model converges to minimize prediction errors across the dataset.

In the Sampler-QNN, a quantum circuit with rotation gates is constructed to process the input data in a quantum manner. A sampler layer is then introduced, which leverages the quantum measurement process to extract expectation values, effectively capturing quantum-enhanced information from the data. During the forward pass, the classical output is passed into the quantum sampler, allowing the quantum layer to refine the predictions made by the classical network. To optimize this process, the Adam optimizer is used to minimize MSELoss, ensuring that the model's predictions are as accurate as possible.

The Variational Quantum Regressor (VQR) relies on a parametric quantum circuit, where trainable parameters are optimized to improve model predictions. Classical layers are defined to complement the quantum circuit, processing the input data before it is passed through the quantum layers. During the forward pass, the VQR applies the parametric quantum circuit to generate predictions based on the input data. Like the other models, 5-fold cross-validation is used to validate the model's robustness, and the Adam optimizer is employed to minimize MSELoss, optimizing the model's parameters to ensure accurate predictions.

The QML with JAX Optimization model combines the power of JAX for differentiable programming and quantum computing via QNodes for feature mapping. The quantum circuit processes the input data, embedding it using quantum gates, and feeds it into a classical-quantum hybrid model. JAX is then used for optimization, where the loss and gradient are defined, allowing for the computation of derivatives to optimize the quantum model. Optax is utilized to minimize the error in predictions, while training is conducted for several epochs to ensure convergence. The model is evaluated using various performance metrics such as R², MAE, RMSE, and MSE to gauge its effectiveness.

\begin{figure}[H]
    \centering
    \includegraphics[width=1\textwidth]{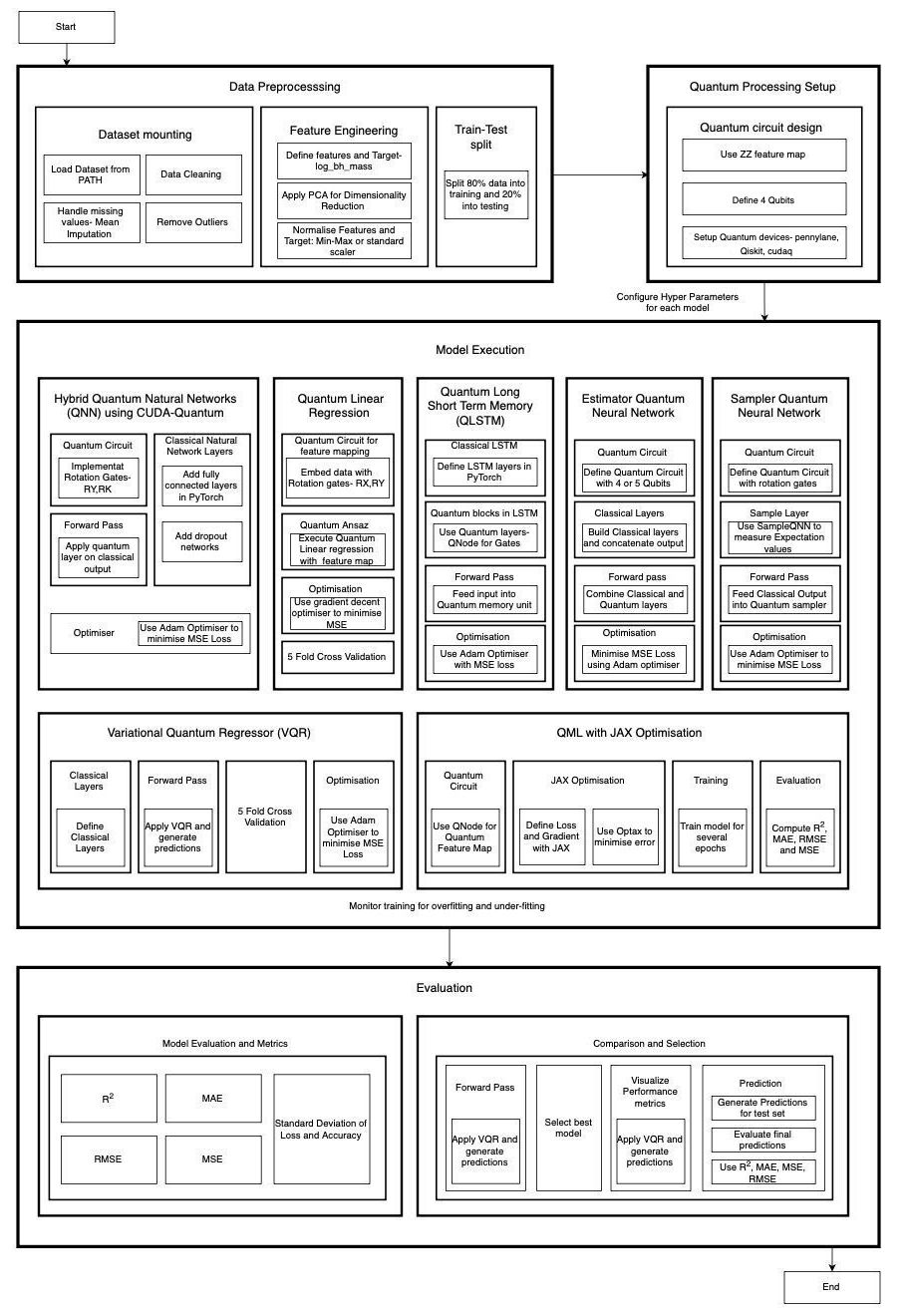}
    \caption{Workflow diagram for Executing QML Algorithms}
    \label{fig:sample_imagefafoawf}
\end{figure}

Fig.\ref{fig:sample_imagefafoawf} describes the workflow for executing the QML algorithms for Blackhole mass Estimation.


\subsection{Comparison of Models and Selection of the Best Model}\label{subsec4}

For this work, classical and quantum ML were compared for their ability to estimate the black hole mass of Type-2 AGN, with many different metrics used to evaluate things like R², MSE, RMSE, and MAE.
Classical models included LR, XGBoost Regression, Random Forest Regressor, `Support Vector Regressor', Lasso Regression, Ridge Regression, Elastic Net Regression, Bayesian Regression, Decision Tree Regressor, Gradient Booster Regressor, Classical Neural Networks, GRU, LSTM, ResNets, and Transformer-Based Regression. Quantum models included Hybrid Quantum Neural Networks, Q-LSTM, Sampler-QNN, Estimator-QNN, VQR, Q-LR, and QML with JAX optimization. All models were trained and tested on the preprocessed data, with different performance metrics provided and compared against the line of best fit.

Once all models were compared, the model with the lowest error was saved across all the specified metrics.

\section{Results}\label{sec5}
Here, we present the results obtained from running various models on the dataset. We have implemented classical ML, DL, and QML algorithms, and analyzed their performance metrics and corresponding accuracies.

\subsection{Classical Machine Learning Algorithms}

\begin{table}[h]
\caption{Performance Metrics for Various Classical Machine Learning Algorithms}
\label{tab:classical_ml}
\centering
\begin{tabular}{@{}lcccc@{}}
\toprule
\textbf{Classical ML Algorithms} & \textbf{MSE} & \textbf{MAE} & \textbf{RMSE} & \textbf{R²} \\ 
\midrule
Linear Regression               & 0.2348       & 0.3921       & 0.4845        & 0.4201 \\ 
XGBoost                         & 0.073        & 0.1837       & 0.2702        & \textbf{0.8196} \\ 
Random Forest Regressor          & 0.0482       & 0.1653       & 0.2195        & \textbf{0.8878} \\ 
SVR                             & 0.1219       & 0.2678       & 0.3491        & 0.7163 \\ 
Lasso Regression                & 0.2773       & 0.4263       & 0.5266        & 0.3152 \\ 
Ridge Regression                & 0.2345       & 0.3919       & 0.4842        & 0.4206 \\ 
Elastic Net Regression          & 0.4052       & 0.4954       & 0.6365        & \textbf{0.0001} \\ 
Bayesian Regression             & 0.4403       & 0.3874       & 0.6069        & 0.0774 \\ 
Decision Tree Regressor         & 0.201        & 0.3319       & 0.4483        & 0.5035 \\ 
Gradient Booster Regressor      & 0.0865       & 0.2212       & 0.294         & 0.7865 \\ 
\botrule
\end{tabular}
\end{table}

\begin{table}[h!]
\caption{Accuracy Metrics for Various Classical Machine Learning Algorithms}
\label{tab:accuracy_ml}
\centering
\begin{tabular}{@{}lccc@{}}
\toprule
\textbf{Algorithm} & \textbf{Accuracy (MSE) \%} & \textbf{Accuracy (RMSE) \%} & \textbf{Accuracy (MAE) \%} \\ 
\midrule
Linear Regression               & 94.32 & 88.28 & 90.52 \\ 
XGBoost                         & \textbf{98.23} & 93.46 & 95.55 \\ 
Random Forest Regressor         & \textbf{98.83} & 94.69 & 96.00 \\ 
SVR                             & 97.05 & 91.56 & 93.52 \\ 
Lasso Regression                & 94.48 & 89.54 & 91.53 \\ 
Ridge Regression                & 94.32 & 88.28 & 90.51 \\ 
Elastic Net Regression          & 91.96 & 87.38 & 90.18 \\ 
Bayesian Regression             & 90.20 & 86.02 & 90.84 \\ 
Decision Tree Regressor         & 95.14 & 89.16 & 91.97 \\ 
Gradient Booster Regressor      & 97.95 & 93.04 & 94.76 \\ 
\bottomrule
\end{tabular}
\end{table}

\begin{figure}[H]
  \centering
  \begin{minipage}{\textwidth}
    \centering
    \includegraphics[width=0.95\linewidth]{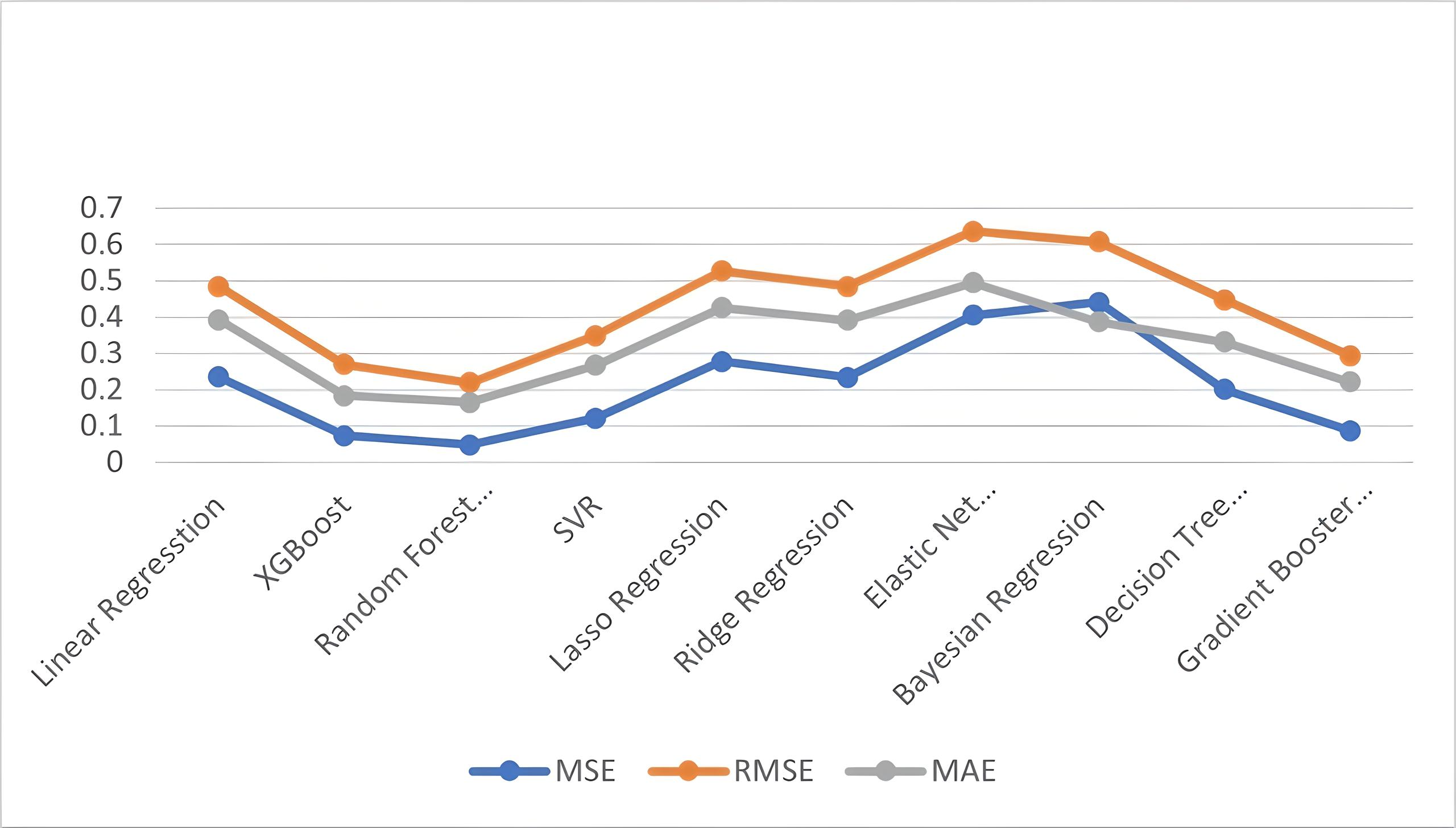}
    \caption{RMSE, MSE and MAE values for Classical Machine Learning Algorithms}
    \label{snjosiadnvoisngvonw}
  \end{minipage}
\end{figure}

\begin{figure}[H]
  \begin{minipage}{\textwidth}
    \centering
    \includegraphics[width=0.95\linewidth]{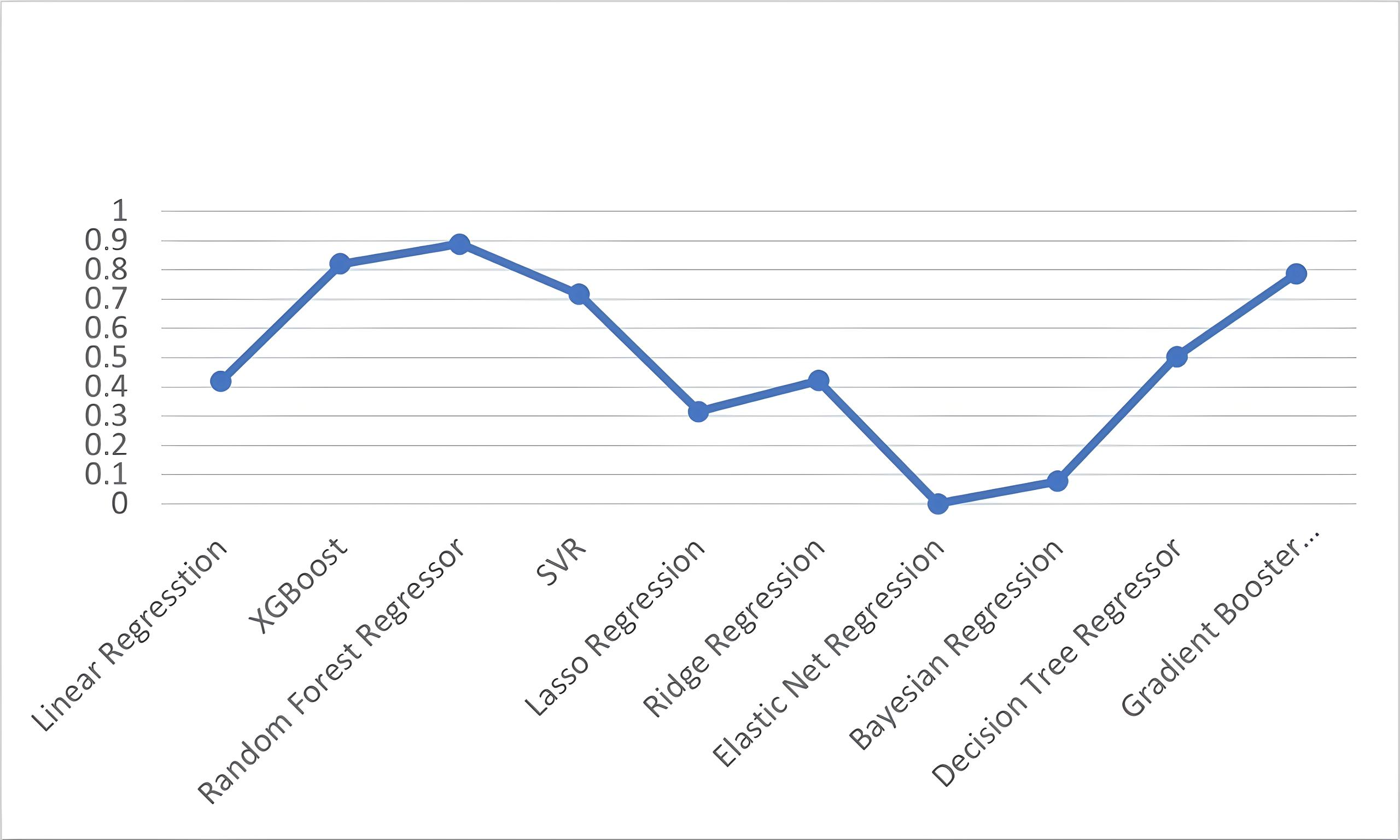}
    \caption{$R^2$ values for Classical Machine Learning Algorithms}
    \label{ssefjisengoerngoenrgonegerngn}
  \end{minipage}

  \vspace{1em}
  
  \begin{minipage}{\textwidth}
    \centering
    \includegraphics[width=0.95\linewidth]{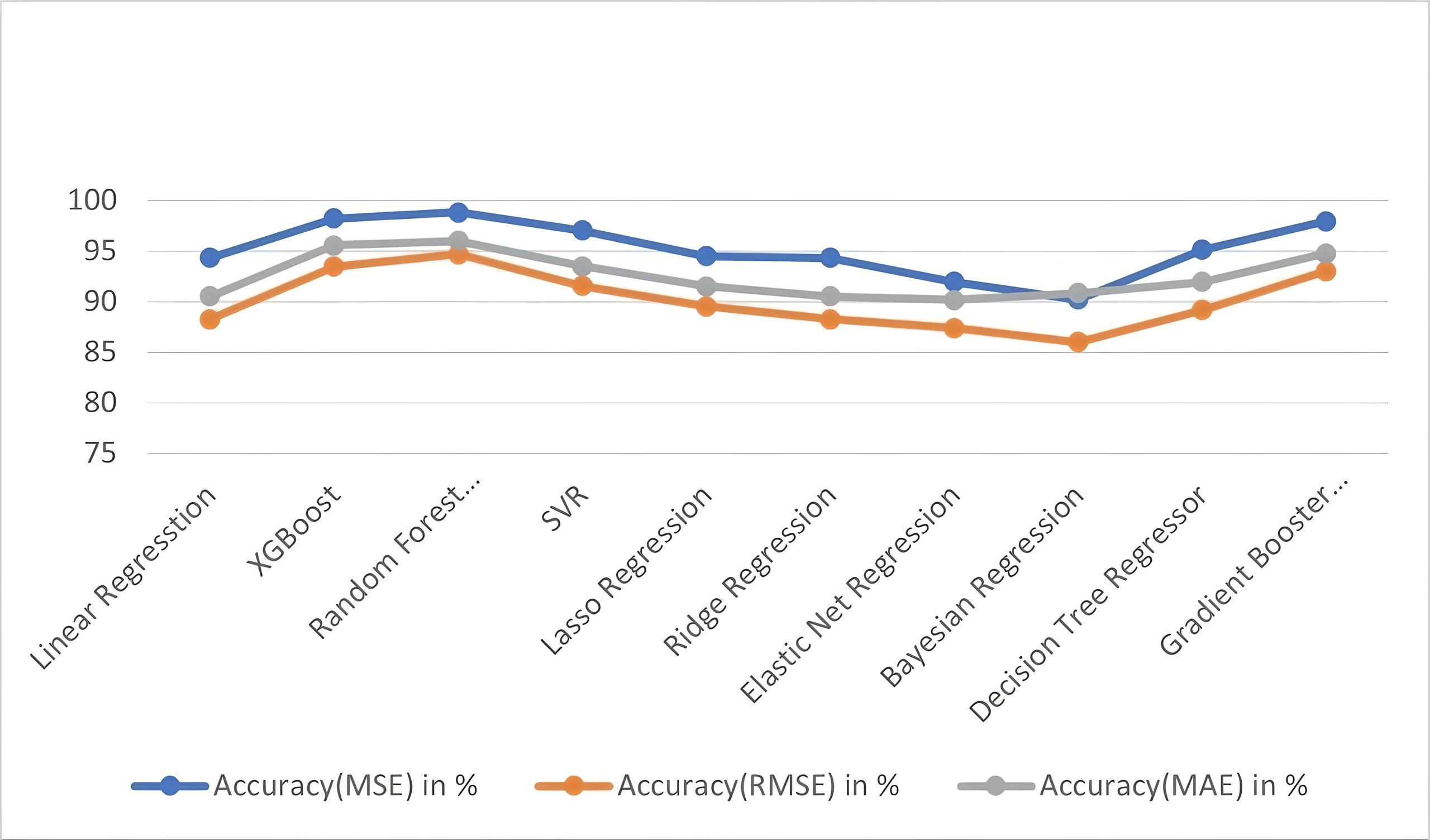}
    \caption{Accuracy based on error metrics (RMSE, MSE, and MAE) for Classical Machine Learning Algorithms}
    \label{aroinoierngohnerhonerohneornhon}
  \end{minipage}
\end{figure}

For evaluating our classical machine learning algorithm performances we used standard metrics like Root Mean Squared Error (RMSE), Mean Square Error (MSE), Mean Absolute Error (MAE), and R-squared. Tables \ref{tab:classical_ml} and \ref{tab:accuracy_ml} and Figures \ref{snjosiadnvoisngvonw}, \ref{ssefjisengoerngoenrgonegerngn}, and \ref{aroinoierngohnerhonerohneornhon} showcase models with MSE-based accuracy ranging from 90.2\% to 98.83\%, with the highest accuracy of 98.83\% belonging to one of the top-performing models, likely the Random Forest Regressor or XGBoost, as indicated by the prior values. These models exhibit robust predictive capability with lower errors and correspondingly higher accuracy rates. Notably, RMSE and MAE also reflect these models' precision. XGBoost and Random Forest Regressor both present high RMSE and MAE accuracies in the range of 91.56\% to 94.69\% and 93.52\% to 96\%, respectively, confirming their ability to provide reliable forecasts with minimal deviation.

The R² values provided in the earlier table further corroborate these findings. Random Forest Regressor, with an R² of 0.8878, and XGBoost, with an R² of 0.8196, demonstrate that these models explain the vast majority of the variance in the data. Their high accuracy percentages across MSE, RMSE, and MAE metrics are a direct reflection of their ability to reduce prediction errors significantly. By contrast, models with lower R² values, such as Elastic Net Regression (0.0001), exhibit lower accuracy percentages, indicating a poorer fit. Thus, combining both error metrics and R² values provides a comprehensive evaluation of each model’s performance, where Random Forest and XGBoost consistently rank highest across these metrics.

  \vspace{1em}

\subsection{Classical Deep Learning Algorithms}

\begin{table}[h]
\caption{Performance Metrics for Various Classical Deep Learning Algorithms}
\label{tab:deep_learning_algorithms}
\centering
\begin{tabular}{@{}lcccc@{}}
\toprule
\textbf{Classical Deep Learning Algorithms} & \textbf{MSE} & \textbf{MAE} & \textbf{RMSE} & \textbf{R²} \\ 
\midrule
Classical Neural Networks    & 0.0787   & 0.1654   & 0.2806   & 0.8055 \\ 
GRU (Gated Recurrent Unit)    & 0.0897   & 0.1876   & 0.2995   & 0.7784 \\ 
Deep Residual Networks (ResNets)  & 0.0931   & 0.2092   & 0.3051   & 0.7700 \\ 
Transformer Based Regression  & 0.0816   & 0.2054   & 0.2856   & 0.7986 \\ 
LSTM                         & \textbf{0.1968 }  & \textbf{0.3448}   & \textbf{0.4433}   & \textbf{0.5144} \\ 
\botrule
\end{tabular}
\end{table}

\begin{table}[h!]
\caption{Accuracy Metrics for Various Classical Deep Learning Algorithms}
\label{tab:accuracy_deep_learning}
\centering
\begin{tabular}{@{}lccc@{}}
\toprule
\textbf{Algorithm} & \textbf{Accuracy (MSE) \%} & \textbf{Accuracy (RMSE) \%} & \textbf{Accuracy (MAE) \%} \\ 
\midrule
Classical Neural Networks      & 98.10 & 93.21 & 96.00 \\ 
GRU (Gated Recurrent Unit)     & 97.83 & 92.76 & 95.46 \\ 
Deep Residual Networks (ResNets) & 98.15 & 93.95 & 95.85 \\ 
Transformer-Based Regression   & 98.38 & 94.34 & 95.93 \\ 
LSTM                           & \textbf{99.71} & \textbf{99.67} & \textbf{99.77} \\ 
\bottomrule
\end{tabular}
\end{table}

\begin{figure}[H]
  \centering
  \includegraphics[width=0.95\linewidth]{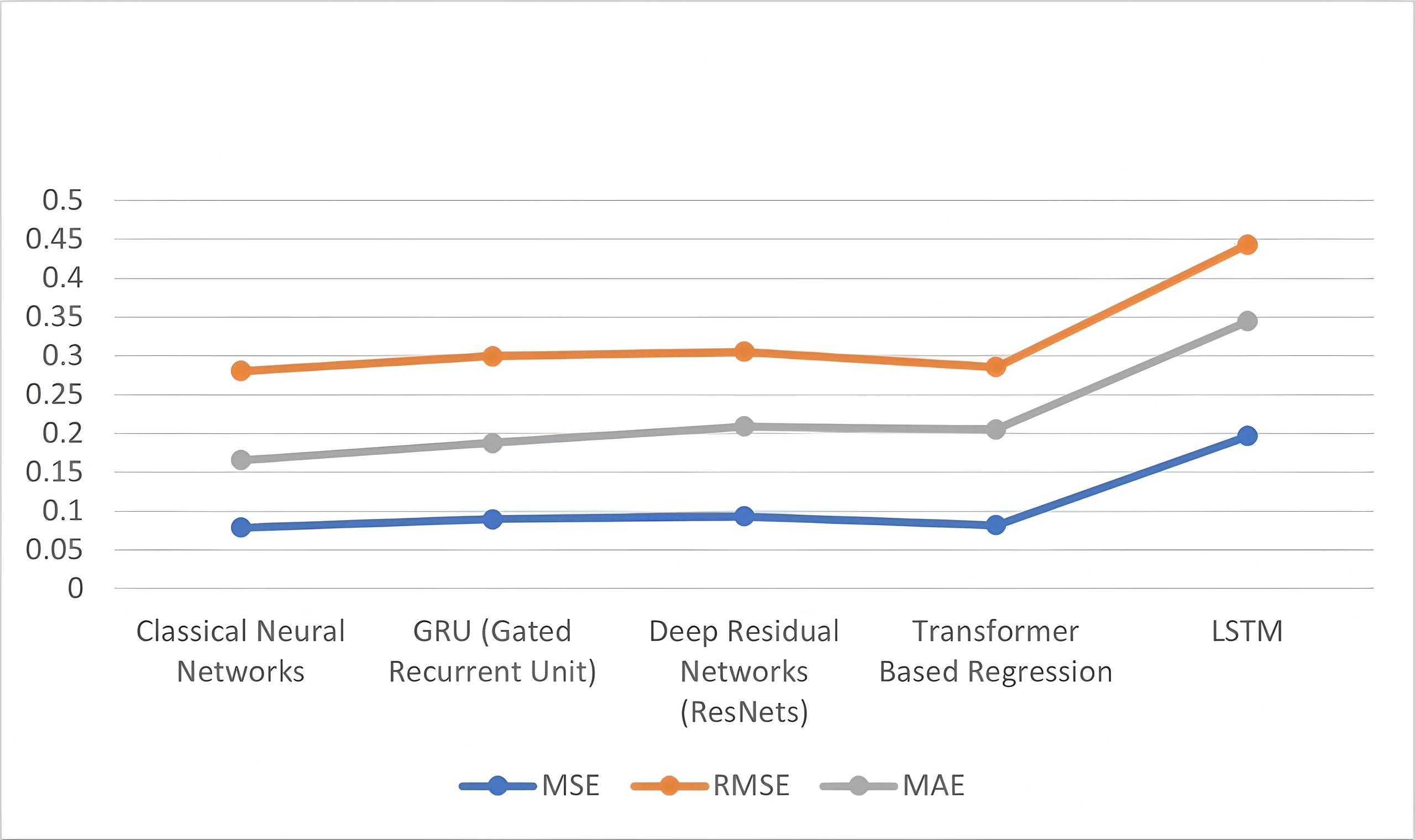}
  \caption{RMSE, MSE and MAE values for Classical Deep Learning Algorithms}
  \label{fig:labelname1dfjeofj}
\end{figure}

\begin{figure}[H]
  \centering
  \includegraphics[width=0.85\linewidth]{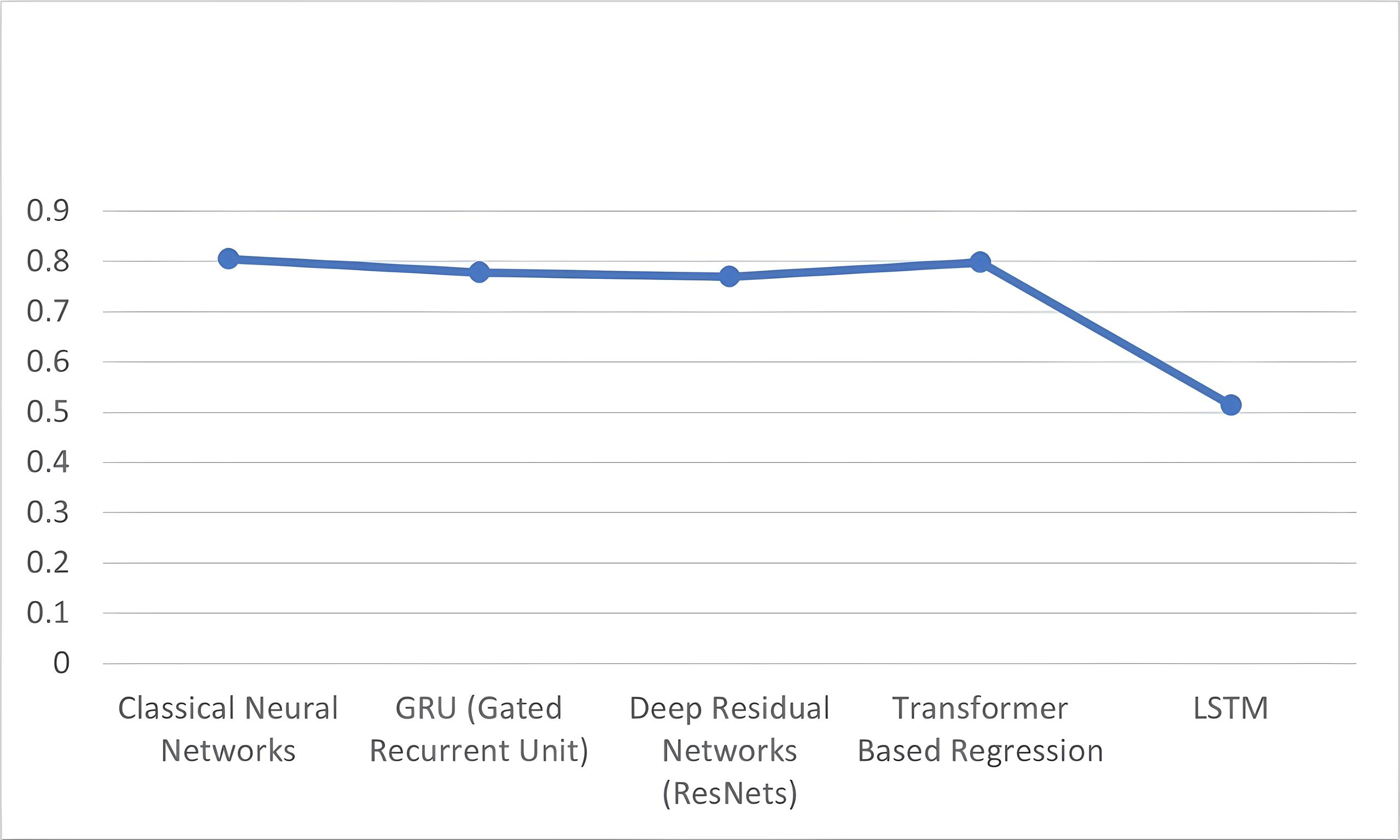}
  \caption{$R^2$ values for Classical Deep Learning Algorithms}
  \label{fig:labelname2asfasegegweg}
  
  \vspace{1em} 
  
  \includegraphics[width=0.85\linewidth]{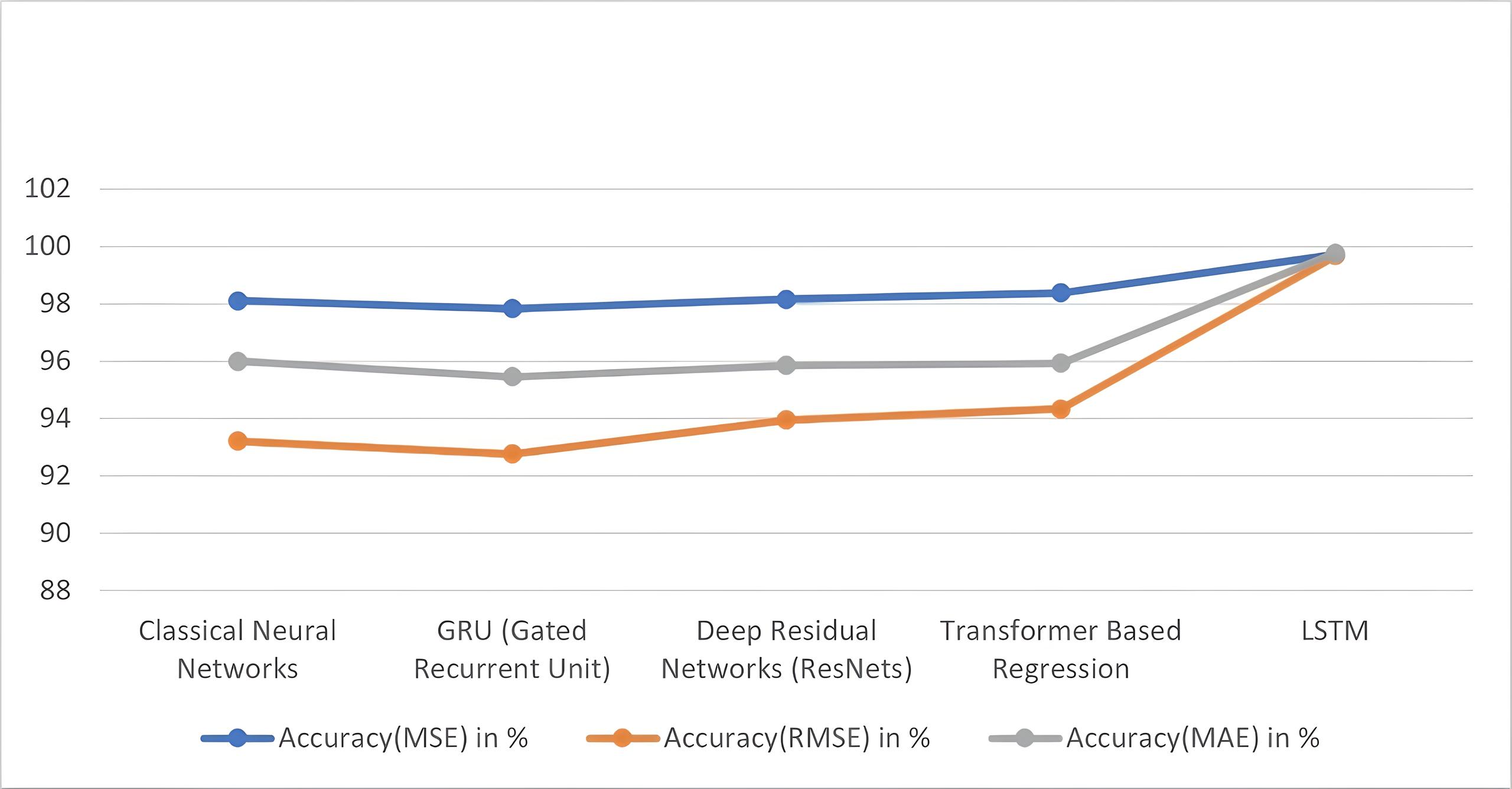}
  \caption{Accuracy based on error metrics (RMSE, MSE, and MAE) for Classical Deep Learning Algorithms}
  \label{fig:labelname3asgsrgrgrg}
\end{figure}

In classical deep learning algorithms, performance metrics such as MSE, RMSE, MAE, and R² comprehensively infer model accuracy and error. Tables~\ref{tab:deep_learning_algorithms} and~\ref{tab:accuracy_deep_learning} and Figures \ref{fig:labelname1dfjeofj}, \ref{fig:labelname2asfasegegweg}, and \ref{fig:labelname3asgsrgrgrg}, visually indicate that the LSTM model achieves the lowest MSE (0.1968), RMSE (0.4433), and MAE (0.3448) but notably has the highest R² value of 0.5144, which signifies that it explains only about 51.44\% of the variance. Interestingly, the LSTM's accuracy metrics based on MSE, RMSE, and MAE are exceptionally high, with MSE accuracy reaching 99.71\%, RMSE accuracy at 99.67\%, and MAE accuracy at 99.77\%, demonstrating its strong predictive power despite the lower R² value. This suggests that while LSTM may not capture variance as effectively as some other models, it consistently delivers highly accurate predictions based on error metrics.

Comparing other models, Classical Neural Networks and Transformer-based Regression perform similarly, with Transformer-based Regression achieving a slightly higher accuracy of 98.38\% in MSE and 94.34\% in RMSE. The GRU model, with a slightly lower R² value of 0.7784, maintains MSE and RMSE accuracies of 97.83\% and 92.76\%, respectively, demonstrating reliable performance. ResNets also perform well, with a high accuracy of 98.15\% in MSE and 93.95\% in RMSE, making them robust options for deep learning tasks. In summary, while the LSTM excels in accuracy metrics, models like ResNets and Transformers offer better performance in explaining variance and reducing errors.

\subsection{Quantum-ML Algorithms}

\begin{table}[h]
\caption{Performance Metrics for Various Quantum Models}\label{tab:quantum_models}
\label{modelperfquantum}
\centering
\begin{tabular}{@{}lcccc@{}}
\toprule
\textbf{Model} & \textbf{MSE} & \textbf{MAE} & \textbf{RMSE} & \textbf{R²} \\ 
\midrule
Hybrid-QNN          & 0.998572    & 0.758497477  & 0.999286     & 0.017 \\ 
Q-LSTM             & 0.322818    & 0.449080192  & 0.56817      & 0.23  \\ 
Sampler-QNN          & 0.0159      & 0.0979       & 0.1262       & 0.34  \\ 
Estimator-QNN      & \textbf{0.025427}    & \textbf{0.012472427}  & \textbf{0.021247}     & \textbf{0.59}  \\ 
VQR                & 0.112692    & 0.269554689  & 0.334998     & 0.28  \\ 
Q-LR                & 0.2301      & 0.4628       & 0.4797       & 0.26  \\ 
QML with Jax Optimization       & 0.3715      & 0.4817       & 0.6095       & 0.0824 \\ 
\botrule
\end{tabular}
\end{table}

\begin{table}[h]
\caption{Accuracy Metrics for Various Quantum Models corresponding to the Error Metrics}\label{tab:quantum_models_accuracies}
\label{quantumaccuracy}
\centering
\begin{tabular}{@{}lccc@{}}
\toprule
\textbf{Model}            & \textbf{RMSE Accuracy (\%)} & \textbf{MSE Accuracy (\%)} & \textbf{MAE Accuracy (\%)} \\ 
\midrule
Hybrid-QNN                 & 87.38                     & 87.39                     & 90.42                     \\ 
Q-LSTM                    & 88.73                     & 93.59                     & 91.09                     \\ 
Sampler-QNN                & 87.39                       & 87.39                       & 90.42                       \\ 
Estimator-QNN             & \textbf{98.76}                       & \textbf{99.75}                       & \textbf{99.18}                       \\ 
VQR                       & 27.6                        & 51.01                       & 41.74                       \\ 
Q-LR                       & 52.03                       & 76.99                       & 53.72                       \\ 
QML with Jax Optimization & 87.91                       & 92.63                       & 90.45                       \\ 
\botrule
\end{tabular}
\end{table}

\begin{figure}[H]
  \centering
  \includegraphics[width=0.95\linewidth]{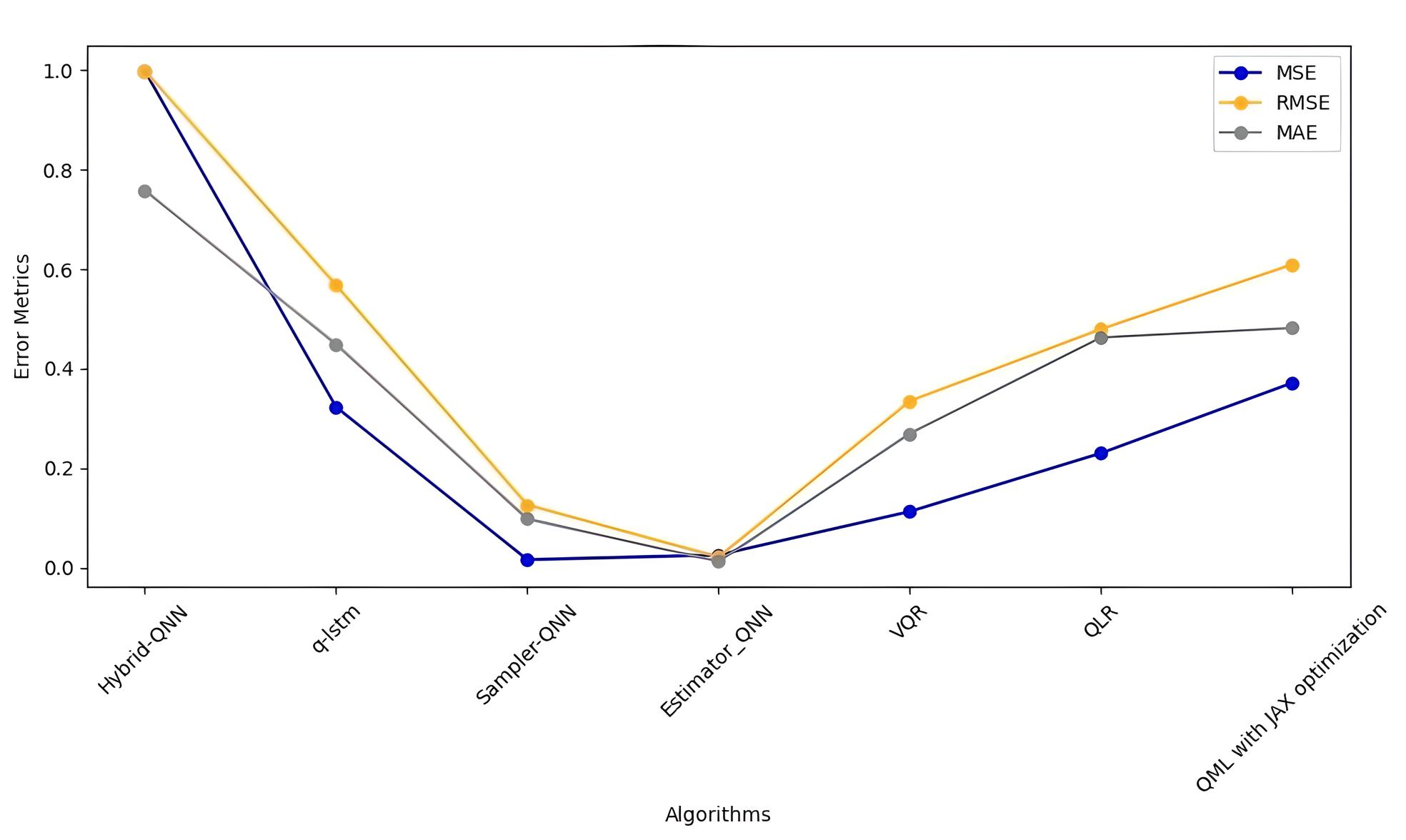}
  \caption{RMSE, MSE and MAE values for Quantum Machine Learning Algorithms}
  \label{fig:labelname1quantum}
  \vspace{1em} 
  \includegraphics[width=0.95\linewidth]{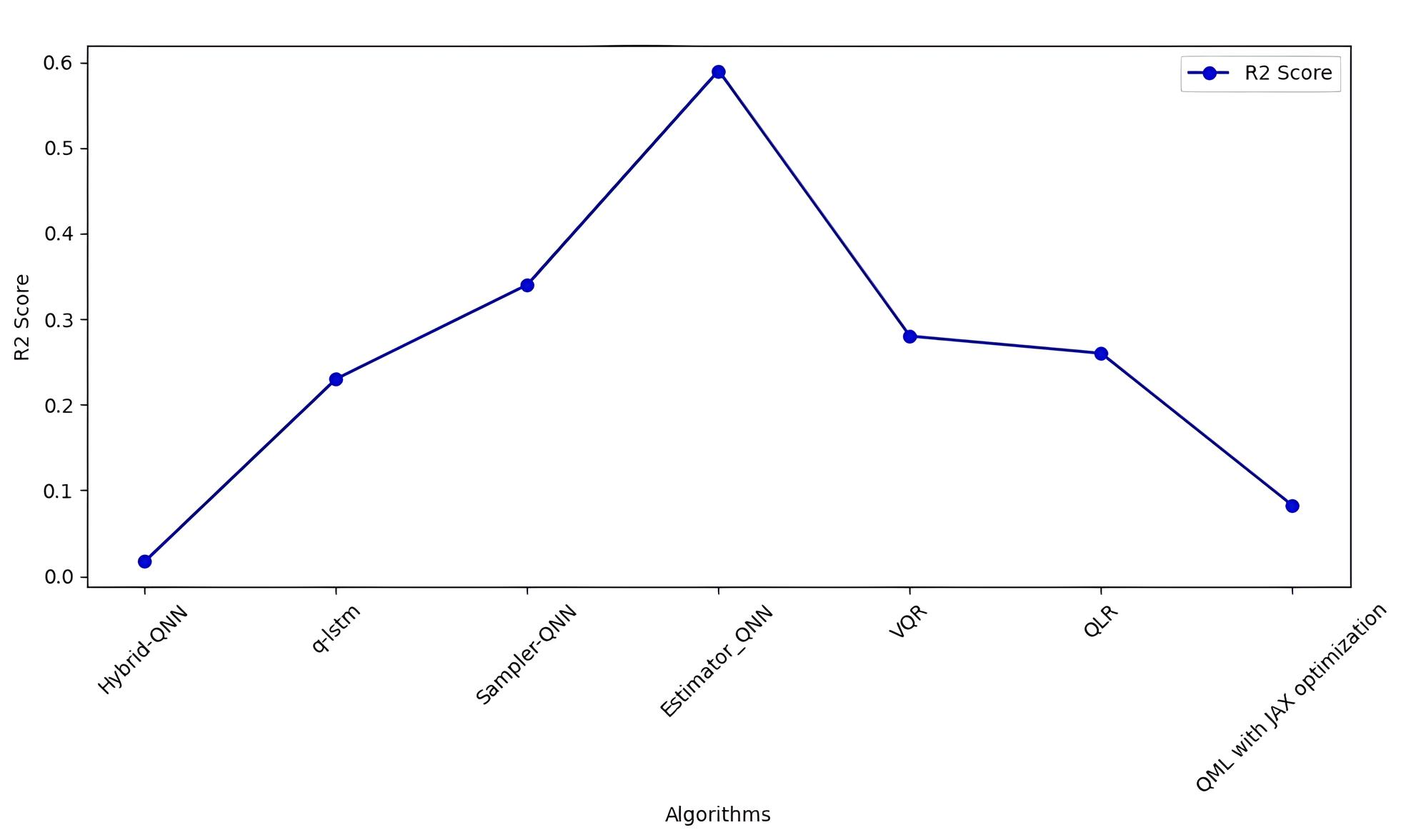}
  \caption{$R^2$ values for Quantum Machine Learning Algorithms}
  \label{fig:labelname2quantum}
\end{figure}

\begin{figure}[H]
  \includegraphics[width=0.95\linewidth]{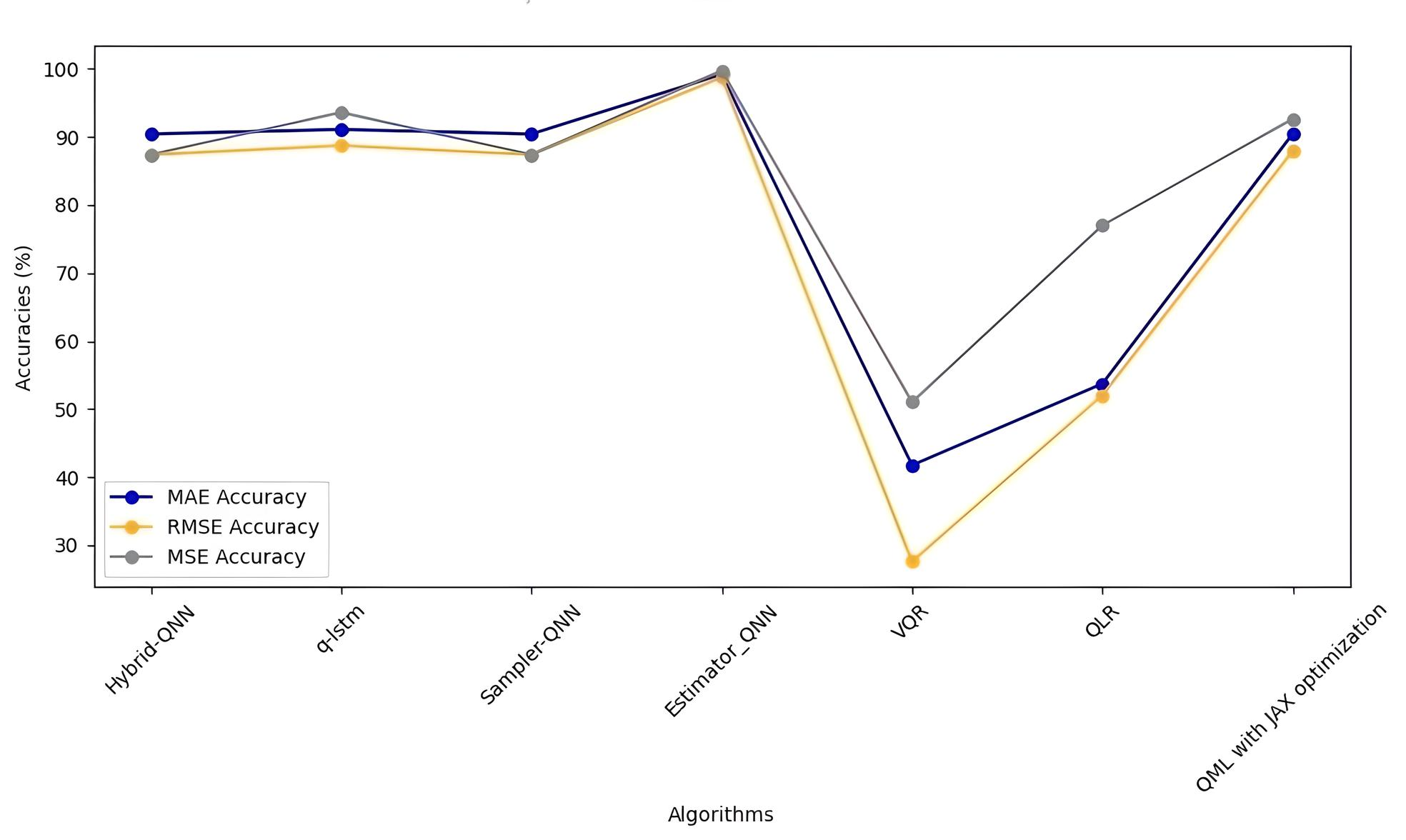}
  \caption{Accuracy based on error metrics (RMSE, MSE, and MAE) for Quantum Machine Learning Algorithms}
  \label{fig:labelname3quantum}
\end{figure}

Tables~\ref{modelperfquantum} and \ref{quantumaccuracy} and Figures~\ref{fig:labelname1quantum}, \ref{fig:labelname2quantum}, and \ref{fig:labelname3quantum} summarize the performance metrics for various quantum models, including MAE, RMSE, MSE, and R² scores. Estimator-QNN demonstrates the best performance with the lowest MAE (0.01247), RMSE (0.021247), and MSE (0.025427), along with the highest R² score of 0.59, indicating a high degree of correlation between predicted and actual values. The Sampler-QNN also performs well with a low MAE (0.0979), RMSE (0.1262), and MSE (0.0159), achieving an R² score of 0.34. Models such as VQR and QLR show relatively higher error values, especially Hybrid-QNN, which has the highest MAE (0.7585) and RMSE (0.9993), resulting in the lowest R² score (0.017). Q-LSTM offers a balanced performance with moderate values across the metrics.

The corresponding accuracy percentages for RMSE, MSE, and MAE provide further insights into the error minimization capabilities of these models. Estimator-QNN excels, achieving an RMSE accuracy of 98.76\%, MSE accuracy of 99.75\%, and MAE accuracy of 99.18\%, reflecting its superior error reduction. Q-LSTM also shows strong accuracy across these metrics, particularly with an MSE accuracy of 93.60\% and an RMSE accuracy of 88.73\%. Hybrid-QNN, despite its high error values, shows decent accuracy in MSE (87.39\%) and RMSE (87.39\%). However, models like VQR and QLR have significantly lower accuracies, indicating their limitations in minimizing errors, with VQR showing the lowest performance across all accuracy percentages. QML with JAX optimization performs relatively well with RMSE and MAE accuracies above 87\%, though its MSE accuracy is slightly lower at 92.63\%.

\section{Discussion}\label{sec6}

In this study, we compared the performance of classical ML algorithms with QML models across a series of standard evaluation metrics, including MSE, RMSE, MAE, and R². Classical algorithms such as XGBoost and Random Forest Regressor stood out with high performance, particularly in MSE-based accuracy, which ranged from 90.2\% to 98.83\%. XGBoost achieved one of the highest accuracies with MSE accuracy at 98.8\%, RMSE accuracy in the range of 91.56\% to 94.69\%, and MAE accuracy at 93.52\% to 96\%. Random Forest performed similarly well, demonstrating robust predictive power with minimal error. These models showed strong generalization capabilities, evidenced by their high R² values, indicating that they captured most of the variance in the data.

Among the classical deep learning models, LSTM emerged as the best-performing model, exhibiting the lowest error values (MSE: 0.1968, RMSE: 0.4433, MAE: 0.3448) while maintaining a high degree of accuracy. Despite its relatively low R² value of 0.5144, the LSTM achieved exceptional accuracy percentages with MSE, RMSE, and MAE accuracies reaching 99.71\%, 99.67\%, and 99.77\%, respectively. This implies that, while the LSTM model may not explain the entire variance within the dataset, it performs well in terms of error minimization, delivering reliable and precise predictions. Other classical deep learning models, such as Transformer-based Regression and ResNets, also performed well, with accuracy metrics close to LSTM but with higher R² values, suggesting that these models may better capture variance while maintaining competitive predictive power.

In contrast, the quantum machine learning models presented more variable performance. Estimator-QNN outperformed other QML models, with the lowest MAE (0.01247), RMSE (0.021247), and MSE (0.025427) alongside the highest R² score of 0.59. This suggests that Estimator-QNN is the most capable of explaining variance and minimizing prediction errors among the quantum models. Sampler-QNN followed closely, performing well with low error values (MAE: 0.0979, RMSE: 0.1262, MSE: 0.0159), though its R² score (0.34) indicated a weaker ability to explain variance compared to Estimator-QNN. On the other hand, models such as Hybrid-QNN struggled with significantly higher error metrics and the lowest R² score of 0.017, reflecting poor predictive capabilities. Q-LSTM showed balanced performance across all metrics, indicating that while QML models hold promise, their performance is not yet on par with classical approaches, especially in variance explanation.

A critical takeaway from the comparison between classical and quantum algorithms is the difference in error minimization and generalization capabilities. Classical models like LSTM, Random Forest, and XGBoost excel at providing highly accurate predictions with minimal error, benefiting from mature optimization techniques and larger datasets. In contrast, quantum algorithms, while showing potential in error reduction, especially with Estimator-QNN, face challenges in scaling and generalization due to quantum noise, limited qubit resources, and less refined optimization techniques. The lower R² values in quantum models suggest that they currently struggle to capture the full complexity of the data. As {\em quantum computing advances, with improvements in qubit stability and error correction, QML models may close the performance gap with classical approaches}. However, at this stage, classical algorithms, particularly LSTM, remain the superior choice for tasks requiring high accuracy and robust predictive power.

LSTM outperforms quantum models due to its mature and well-optimized architecture, which has been refined through years of research and practical application. Designed to handle sequential data, LSTM’s memory cells and gates allow it to capture long-term dependencies effectively. It benefits from advanced optimization techniques like Adam and RMSprop, leading to stable training and better convergence, especially on large datasets. Unlike quantum models, LSTM does not face challenges with data encoding and qubit limitations, allowing it to scale efficiently using modern GPUs and TPUs. LSTM’s ability to generalize from noisy or incomplete data and its robustness across various domains make it more practical for real-world applications. Meanwhile, quantum models struggle with noise, limited qubit resources, and optimization inefficiencies, making them less competitive in terms of generalization and performance at this stage.

In our study, we rely on quantum simulations rather than real quantum hardware to explore the performance of QML algorithms. Unlike real quantum computers, simulations do not introduce quantum noise or decoherence, which are critical challenges in physical quantum systems. Quantum noise arises from external environmental interactions and imperfect gate operations, leading to errors and reduced algorithmic performance, while decoherence causes the loss of quantum information. Since these effects are absent in simulations, the results appear idealized and do not fully reflect the behavior of quantum algorithms in real-world conditions. However, simulations allow for precise testing and scalability of QML models without the constraints of current noisy intermediate-scale quantum (NISQ) devices. Despite these advantages, simulations have their limitations, such as lacking real-time error behavior and quantum parallelism, and they are constrained by classical computational resources, which limits the exploration of larger quantum systems. Thus, while simulations provide valuable insights, they may not capture the full complexity of executing QML on actual quantum hardware, which is critical to account for in future work.

Current quantum processors have a limited number of qubits, restricting the size and complexity of QML models that can be implemented, making many algorithms impractical for real-world applications. Additionally, QML algorithms face optimization challenges in noisy environments, where quantum gate noise and imperfect measurements can result in sub-optimal convergence. Another significant limitation is the inefficiency of encoding classical data into quantum states, especially for large datasets, which can hinder the realization of the quantum speed-up that QML promises.

To improve QML, several advancements are necessary, including increasing qubit count, enhancing qubit stability, reducing noise, and implementing more efficient error correction methods. Scalable quantum circuits, better data encoding, improved generalization, and the development of advanced quantum neural network architectures are essential for enhancing QML performance. Hybrid quantum-classical models offer potential by combining quantum speedup with classical stability, but they come with trade-offs. While hybrid models allow quantum processors to tackle tasks like optimization and pattern recognition, they are limited by the small role quantum components currently play, with classical stability often taking precedence. This adds complexity and practical challenges in integrating quantum and classical systems, especially as quantum devices have limited scalability and remain costly as of now, hybrid models reduce quantum resource demands but introduce overhead and are prone to quantum noise, which can degrade performance. Accessible programming frameworks and reduced resource costs are also needed to make QML more practical. Ultimately, hybrid models hold promise, but optimizing them requires addressing these trade-offs to fully exploit quantum advantages.

\section{Conclusion}\label{sec7}

In conclusion, this comparative analysis of classical and QML algorithms for black hole mass estimation in Type-2 AGNs highlights the clear advantages of classical models in terms of error minimization and generalization. Among the classical approaches, LSTM outperformed all other models with its highly accurate predictions, achieving a remarkable MSE accuracy of 99.71\%, RMSE accuracy of 99.67\%, and MAE accuracy of 99.77\%. Other classical models, such as XGBoost and Random Forest Regressor, also demonstrated strong performance with high R² values and robust accuracy metrics. The consistency and reliability of classical algorithms, along with their ability to handle larger datasets and reduce errors effectively, make them more practical for tasks such as black hole mass estimation. These models benefit from years of research, optimization, and accessible computational resources, making them highly effective for large-scale astrophysical data analysis.

On the other hand, while showing promise, the quantum models are not yet on par with classical algorithms in terms of overall performance. Estimator-QNN was the best-performing quantum model, achieving an MSE accuracy of 99.75\% and demonstrating the ability to minimize errors. However, quantum models, including Q-LSTM and Hybrid-QNN, struggled with lower R² scores and higher error metrics, indicating challenges in generalization and optimization. Limitations such as qubit restrictions and inefficient data encoding have constrained the models' capacity to learn complex patterns from the data. In particular, the process of encoding classical data into quantum states remains inefficient, leading to information loss, which affects model accuracy. Additionally, the limited number of qubits available in quantum simulators further exacerbates these issues by restricting the model's ability to handle high-dimensional data effectively.

Although quantum noise and decoherence are not factors in quantum simulations, the simulated quantum models still face inherent challenges in terms of optimization. Classical models, such as LSTM, have benefited from decades of advancements in optimization algorithms, which enable better convergence and generalization. In contrast, quantum models are still in the early stages of developing efficient optimization techniques. Furthermore, quantum models often require extensive tuning of hyperparameters to achieve comparable performance, adding to their complexity. While quantum hardware and algorithms are rapidly evolving, the gap in performance between classical and quantum machine learning remains significant. As quantum computing technology matures, it is expected that these models will gradually close the performance gap. However, based on the results of this study, classical machine learning models, particularly LSTM, remain the superior choice for black hole mass estimation in Type-2 AGNs at this stage.

\section{Future Scope}\label{sec8}

In the future, exploring QML algorithms for black hole mass estimation in Type-2 AGNs presents significant potential, especially as quantum computing technology matures. While classical algorithms, particularly LSTM, currently outperform quantum approaches in terms of R², MSE, and RMSE, ongoing advancements in quantum hardware and optimization techniques could close this gap. Future research could focus on refining hybrid quantum-classical models, incorporating larger datasets, and addressing quantum noise and error correction challenges. Additionally, exploring other quantum algorithms or integrating more sophisticated quantum circuits could potentially lead to breakthroughs in the accuracy and efficiency of black hole mass predictions. As quantum computing evolves, it may provide novel methods for handling large-scale astrophysical datasets, offering deeper insights into AGN dynamics and black hole-host galaxy co-evolution.

\textbf{Code Availability} The code for this study's implementation is available at:
\noindent\textit{https://github.com/sathwikNARKEDimilli29/BlackHoleMassEstimation}

\section{Ethical Consideration}

In preparing this manuscript, large language models were used for refining grammar and enhancing the overall clarity of the text. Their application was strictly limited to editorial support and did not impact the content, analysis, or interpretation of the research findings.

\section{Funding Declaration}

This research did not receive any specific grant from funding agencies in the public, commercial, or not-for-profit sectors.


\section*{Appendix-1: Evaluation Metrics}

\subsection*{Mean Absolute Error}
The Mean Absolute Error (MAE) is a performance measure that calculates the average of the absolute differences between predicted and actual values. It indicates how far the predictions are from the actual outcomes on average, ignoring the direction of the errors. This metric gives an overall sense of the accuracy of the model.
\[
\boxed{\text{MAE} = \frac{1}{n} \sum_{i=1}^{n} \left| y_i - \hat{y_i} \right|}
\]
Where:
\begin{itemize}
  \item $n$ is the number of observations,
  \item $y_i$ is the actual value for observation $i$,
  \item $\hat{y_i}$ is the predicted value for observation $i$.
\end{itemize}

\subsection*{Mean Squared Error}
The Mean Squared Error (MSE) evaluates the average of the squared differences between the predicted and actual values. By squaring the errors, larger mistakes have a greater impact on the final result, making it a useful metric for highlighting significant prediction deviations. It essentially shows how much the predictions deviate from the actual data.
\[
\boxed{\text{MSE} = \frac{1}{n} \sum_{i=1}^{n} (y_i - \hat{y_i})^2}
\]
Where:
\begin{itemize}
  \item $n$ is the number of observations,
  \item $y_i$ is the actual value for observation $i$,
  \item $\hat{y_i}$ is the predicted value for observation $i$.
\end{itemize}

\subsection*{Root Mean Squared Error}
The Root Mean Squared Error (RMSE) is the square root of the average of the squared differences between the predicted and actual values. This metric provides a measure of error magnitude in the same units as the target variable, and, like MSE, it gives more weight to larger discrepancies, emphasizing bigger errors in the model’s performance.
\[
\boxed{\text{RMSE} = \sqrt{\text{MSE}}}
\]

\subsection*{R-squared}
R-squared ($R^2$) represents the proportion of the total variance in the target variable that is explained by the model's predictions. Ranging between 0 and 1, a higher $R^2$ value suggests a better fit of the model to the observed data, meaning the model explains a greater portion of the variability in the outcomes.
\[
\boxed{R^2 = 1 - \frac{\sum_{i=1}^{n} (y_i - \hat{y_i})^2}{\sum_{i=1}^{n} (y_i - \bar{y})^2}}
\]
Where:
\begin{itemize}
  \item $y_i$ is the actual value for observation $i$,
  \item $\hat{y_i}$ is the predicted value for observation $i$,
  \item $\bar{y}$ is the mean of the actual values,
  \item $n$ is the number of observations.
\end{itemize}

\begin{equation}
\bar{y} = \frac{1}{n} \sum_{i=1}^{n} y_i
\end{equation}
Where:
\begin{itemize}
  \item $\bar{y}$ is the mean of the actual values.
\end{itemize}

\subsection*{Custom Accuracy Metric: Accuracy by Error}

\[
\boxed{
\text{accuracy} (\%) = \left( 1 - \frac{\text{error\_metric}}{\text{range of } y} \right) \times 100
}
\]
Here,
\begin{itemize}
  \item  \( y \) is the target variable.
  \item The error metric includes RMSE, MSE, and MAE.
\end{itemize}

\end{document}